\documentclass[fleqn,12pt,twoside]{article}
\usepackage{espcrc1}
\def\beq{\begin{equation}}
\def\eeq{\end{equation}}

\def\NP{{\it Nucl.~Phys.~}}

\def\PL{{\it Phys.~Lett.~}}

\def\vol#1{{\bf #1}}
\def\vyp#1#2#3{\vol{#1} (#2) #3}

\def\epm#1#2{\hbox{${\lower1pt\hbox{$\scriptstyle +#1$}}
\atop {\raise1pt\hbox{$\scriptstyle -#2$}}$}}

\def\gsim{\mathrel{\rlap{\lower4pt\hbox{\hskip1pt$\sim$}}
    \raise1pt\hbox{$>$}}}         
\def\eg{{\it e.g.}}

\def\frac#1#2{{{#1}\over {#2}}}
\def\half{\hbox{${1\over 2}$}}

\def\as{\alpha_s}

\catcode`@=11 
\def\slash#1{\mathord{\mathpalette\c@ncel#1}}
 \def\c@ncel#1#2{\ooalign{$\hfil#1\mkern1mu/\hfil$\crcr$#1#2$}}
\def\lsim{\mathrel{\mathpalette\@versim<}}
\def\gsim{\mathrel{\mathpalette\@versim>}}
 \def\@versim#1#2{\lower0.2ex\vbox{\baselineskip\z@skip\lineskip\z@skip
       \lineskiplimit\z@\ialign{$\m@th#1\hfil##$\crcr#2\crcr\sim\crcr}}}
\catcode`@=12 

\def\be{\begin{equation}}
\def\ee{\end{equation}}
\def\bea{\begin{eqnarray}}
\def\eea{\end{eqnarray}}

\def\epm#1#2{\hbox{${\lower1pt\hbox{$\scriptstyle +#1$}}
\atop {\raise1pt\hbox{$\scriptstyle -#2$}}$}}
\def\wup{{W^+}}

\usepackage{epsfig,graphicx,wrapfig}
\usepackage[figuresright]{rotating}
\begin{document}

\newcommand{\ttbs}{\char'134}
\newcommand{\AmS}{{\protect\the\textfont2
  A\kern-.1667em\lower.5ex\hbox{M}\kern-.125emS}}

\hyphenation{author another created financial paper re-commend-ed Post-Script}

\setcounter{page}{0}
\pagestyle{empty}
\begin{flushright}
{\tt hep-ph/0207209}\\
{RM3--TH/02-10}\\
\end{flushright}
\vglue.3cm
\begin{center}
{\bf\Large Hadron Physics at a Neutrino Factory}
\vglue.5cm

Stefano Forte
\vglue.3cm

{\it INFN, Sezione di Roma III,\\ via della Vasca Navale 84,
  I--00146 Roma, Italy}
\vglue2.cm
{\bf Abstract}
\end{center}
We review the way intense neutrino
beams at the front--end of a muon storage ring can be used to probe
the structure of  hadrons. 
Specifically, we discuss how the polarized and
unpolarized flavor structure of the nucleon can be
disentangled in inclusive deep-inelastic scattering, and how less
inclusive measurements can shed light on various aspects of hadron
structure such as fragmentation functions or generalized parton distributions.


\vskip 2.5cm
\begin{center}
 Invited talk at
{\bf QCD-N'02}\\
Ferrara, Italy, April 2002\\
{\it to be published in the proceedings}
\end{center}

\newpage
\pagestyle{plain}
\title{Hadron physics at a neutrino factory}

\author{Stefano Forte\address[INFN]{INFN, Sezione di Roma III,\\
via della Vasca Navale 84, I--00146 Roma, Italy}%
 }

\maketitle

\begin{abstract}

We review the way intense neutrino
beams at the front--end of a muon storage ring can be used to probe
the structure of  hadrons. 
Specifically, we discuss how the polarized and
unpolarized flavor structure of the nucleon can be
disentangled in inclusive deep-inelastic scattering, and how less
inclusive measurements can shed light on various aspects of hadron
structure such as fragmentation functions or generalized parton distributions.
\end{abstract}      

\section{PHYSICS AT A NEUTRINO FACTORY}
The use of a muon storage ring as an intense source of neutrinos has
attracted renewed attention: it is  a natural
stepping stone toward a muon collider, and it would offer a wealth of
opportunities for doing physics both of neutrinos (long baseline), and with
neutrinos (front-end). Whereas the physics of neutrino oscillations, and in
particular the possibility of discovering CP violation in the neutrino
sector provide some of the strongest motivations for a neutrino
factory, a sizable part of the physics program at such a facility
should be devoted to the use of the neutrino beam as a probe of matter.

Front--end physics at a neutrino factory is based on the realization
that, because of the flavor and spin structure of the coupling of
neutrinos to weak currents,  a neutrino beam has a greater 
physics potential than conventional electron
or muon beams, and
it can be  used to perform tests of unsurpassed
precision of the standard model, and as a probe of hadron
structure of unique sensitivity.

Here we will briefly review the potential of neutrino beams as probes
of matter. First, we will review deep-inelastic scattering
(DIS) with neutrino beams, and show that it can be used to 
sort out the flavor
structure of parton distributions of the nucleon, both polarized and
unpolarized.
Then, we will consider increasingly less
inclusive measurements, and see how they can be used to shed light
on various other aspects of hadron structure, such as 
fragmentation functions and generalized parton distributions. 

The results
discussed here  are  based on a recent detailed quantitative
study performed by a CERN working group~\cite{cernrep}; quantitative
estimates given here are taken from there unless otherwise stated, and
are based on the `CERN  scenario'~\cite{cernmunu}
for a neutrino factory: 
specifically, a
 50~GeV $\mu$ beam, with $10^{20}$ muon decays per year along a 100~m
straight section. 
For other studies on the physics
potential of neutrino factories see Ref.~\cite{usrep}; a recent status
report on the neutrino factory/muon collider project is in Ref.~\cite{status}.

\section{DEEP-INELASTIC SCATTERING WITH PARITY VIOLATION}

Inclusive DIS is the standard way of accessing
the parton content of hadrons. The use of neutrino beams allows one to
study DIS mediated by the weak, rather than electromagnetic interaction.
 The neutrino-nucleon deep-inelastic
cross section for charged--current interactions, up to
corrections suppressed by powers of $m_p^2/Q^2$ is given by
\bea
\label{disxsec}
&&
\frac{d^2\sigma^{\lambda_p\lambda_\ell}(x,y,Q^2)}{dx dy}
=
\frac{G^2_F}{  2\pi (1+Q^2/m_W^2)^2}
\frac{Q^2}{ xy}\Bigg\{
\left[-\lambda_\ell\, y \left(1-\frac{y}{2}\right) x { F_3(x,Q^2)}
      \right.\nonumber\\ &&\quad\left.
+(1-y) { F_2(x,Q^2)} + y^2 x {
F_1(x,Q^2)}\right]
 -2\lambda_p
  \left[
     -\lambda_\ell\, y (2-y)  x { g_1(x,Q^2)}
\right.\nonumber\\ &&\qquad\left. -(1-y) {
g_4(x,Q^2)}- y^2 x { g_5(x,Q^2)}
  \right]
\Bigg\},
\eea
where $\lambda$ are the lepton and proton helicities (assuming
longitudinal proton polarization), and the
kinematic variables are $y=\frac{p\cdot q}{p\cdot
k}$ (lepton fractional energy loss), $x= \frac{Q^2}{2 p\cdot
q}$ (Bjorken $x$). 
The neutral--current cross--section is found from Eq.~(\ref{disxsec})
by letting
$m_W\to m_Z$ and multiplying by an
overall factor  $[\half(g_V-\lambda_\ell
g_A)]^2$.

The advantage of $W$ and $Z$-mediated DIS over conventional 
$\gamma^*$ DIS  is clear when inspecting the
parton content of the polarized and unpolarized structure functions
$F_i$ and $g_i$. Up to $O(\alpha_s)$ corrections, in terms of the
unpolarized and polarized quark distribution for the $i$--th flavor
 $q_i\equiv
q_i^{\uparrow\uparrow}+q_i^{\uparrow\downarrow}$ and 
$\Delta q_i\equiv 
q_i^{\uparrow\uparrow}-q_i^{\uparrow\downarrow}$
\begin{center}
\begin{tabular}[c]{ccc}

NC&$F_1^{\gamma} =\half\sum_{i}  e^2_i\left(q_i+\bar
q_i\right)$\quad\qquad&$ g_1^{\gamma}=\half\sum_{i}
e^2_i\left(\Delta q_i+\Delta \bar
q_i\right)$\\
NC&$F_1^{Z} =\half\sum_{i}  (g^2_V+g^2_A)_i\left(q_i+\bar
q_i\right)$\quad\qquad&$ g_1^{Z}=\half\sum_{i} (g^2_V+g^2_A)_i
\left(\Delta q_i+\Delta \bar
q_i\right)$\\
NC&$F_3^{Z} =2\sum_{i}  (g_Vg_A)_i \left(q_i+\bar
q_i\right)$\quad\qquad&$ g_5^{Z}=-\sum_{i}(g_Vg_A)_i
\left(\Delta q_i+\Delta \bar
q_i\right)$\\
CC&{ $F_1^\wup =\bar u + d + s + \bar c$}\quad\qquad&${ g_1^\wup=\Delta\bar u + \Delta d +
\Delta s + \Delta \bar c}$\\
CC&${ -F_3^\wup/2 = \bar u - d - s +\bar c }$\quad\qquad & { $g_5^\wup = \Delta \bar u -\Delta d -\Delta s +\Delta\bar c$}\\
\phantom{CC}& $F_2= 2 x F_1$& $g_4= 2 x g_5$\\
\end{tabular}
\end{center}

\begin{figure}[t]
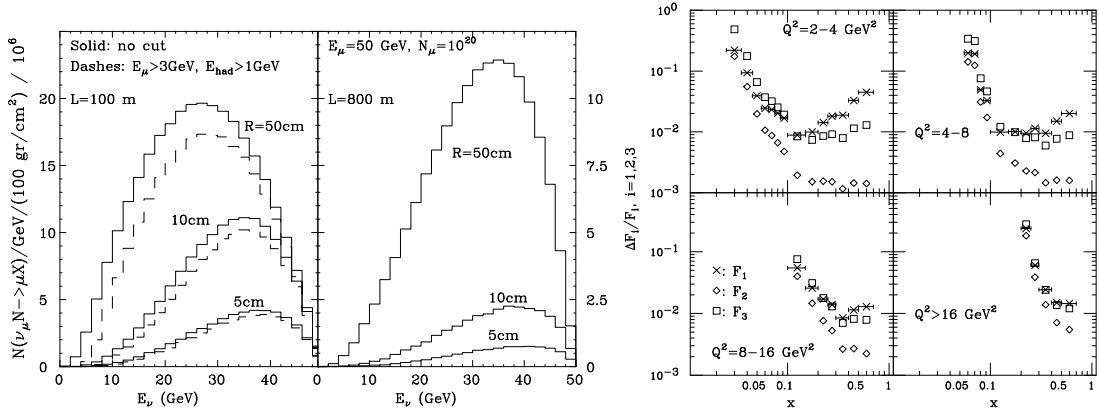

\begin{minipage}[t]{80mm}
\includegraphics[width=19pc]{50gev.eps}
\vskip-.5cm
\end{minipage}
\hspace{\fill}
\begin{minipage}[t]{80mm}
\includegraphics[width=15pc]{F123err.eps}
\vskip-.5cm
\end{minipage}
\vskip-.5cm
\caption{\label{strfun}
Charged-current event rates for several detector
and beam configurations (left), and 
expected errors on the determination of the individual
nucleon structure functions (right) for one year of running.}
\vskip-.6cm\end{figure}
Here $e_i$ are the electric charges and $(g_V)_i$, $(g_A)_i$ are the weak
charges of the $i$--th quark flavor.
If $W^+\to W^-$ (incoming $\bar \nu$ beam), then
$u\leftrightarrow d,\>c\leftrightarrow s$. The structure functions
$F_3$, $g_4$ and $g_5$ are parity--violating, and therefore not
accessible in virtual photon scattering. 
Of course, beyond leading order in the strong coupling each
quark or antiquark flavor's contribution receives $O(\as)$
corrections proportional to itself and to all other quark, antiquark
and gluon distributions. The gluon correction is flavor--blind, and
thus decouples from the parity--violating structure functions $F_3$,
$g_4$ and $g_5$. 

Hence, thanks to the weak couplings, more independent linear
combination of individual quark and antiquark distributions are accessible.
A further advantage of a neutrino factory follows from  the fact that
the neutrino beam has a broad-band energy spectrum (Fig.~\ref{strfun}). 
Because
$y=Q^2/(2 x m_p E_\nu)$, at fixed $x$ and $Q^2$, $y$ 
only varies with the beam energy. Hence, at a neutrino factory it is
possible to disentangle the individual structure functions which make
up the cross section Eq.~(\ref{strfun}) by measuring the neutrino
energy  on an event-by event basis, and then fitting
the $y$ dependence of the data for fixed $x$ and $Q^2$ (Fig.~\ref{strfun})

\subsection{Unpolarized DIS and precision tests of the standard model}
\begin{figure}[t]
\begin{minipage}[t]{70mm}
\includegraphics[width=17pc]{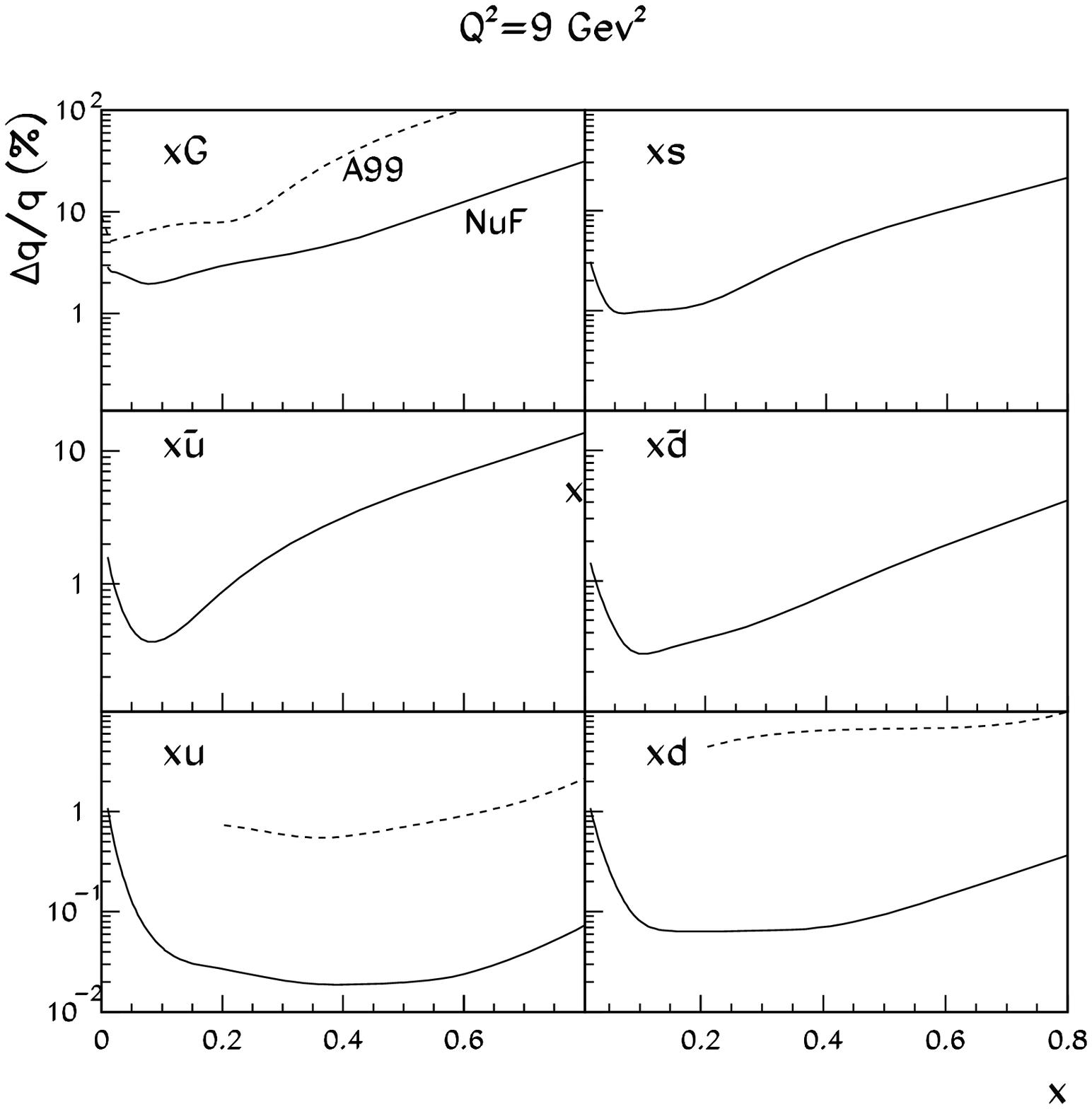}
\vskip-.7cm
\end{minipage}
\hspace{\fill}
\begin{minipage}[t]{85mm}
\includegraphics[width=18pc]{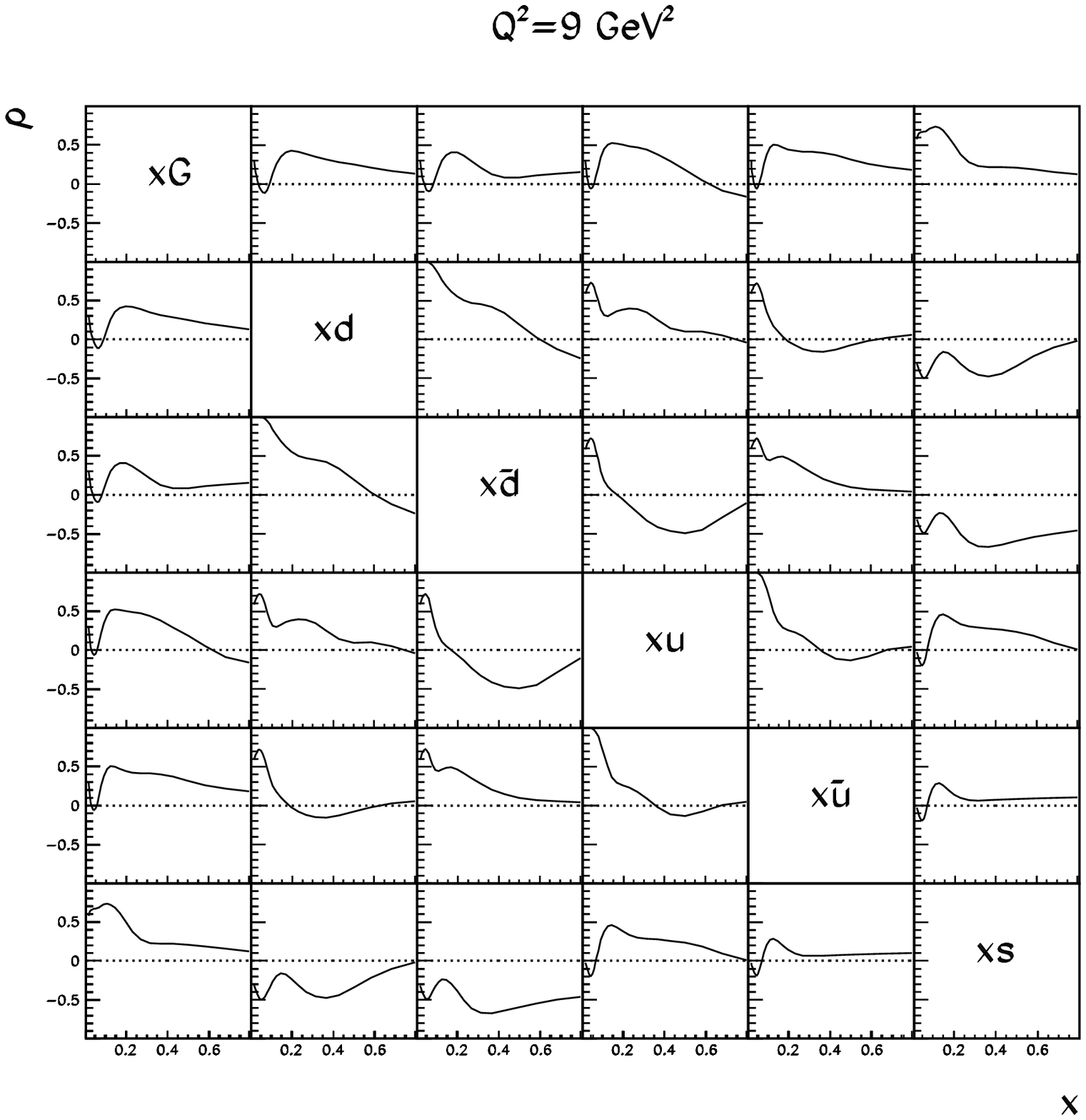}
\vskip-.7cm
\end{minipage}
\vskip-.7cm
\caption{\label{pdfec}
Percentage errors (left, solid) and correlation coefficients
(right) of parton distributions at a $\nu$ factory
compared to
present--day~\cite{alek} errors (left, dashed).
}
\vskip-.65cm\end{figure}

Unpolarized parton distributions are a necessary ingredient in the
computation of any collider process. However, only the up, down and
gluon distributions can be determined in a reasonably accurate way
from present-day DIS data~\cite{pdfs,alek}. Also,
only the combination $q_i+\bar
q_i$ can be extracted from $\gamma^*$-mediated inclusive DIS.
Some information on strangeness can be extracted~\cite{BPZ} 
from neutrino data, while some less--inclusive observables (such as
$W$ production, or Drell--Yan)  provide some constraints on the
relative size of the $q$ and $\bar q$ distributions, but the results
are at best semi--quantitative (see Fig.~\ref{pdfec}).

Thanks to the availability of more independent combinations of parton
distributions, full flavor separation would be possible at a 
neutrino factory.
In Fig.~\ref{pdfec} 
error estimates on individual partons at a neutrino  factory
are compared to the extant knowledge. 
Note that no current errors on strange and
antiquark distributions are given, since the present results largely depend
on theoretical prejudice.
 In Fig.~\ref{pdfec} we further show that the
point--by--point correlation of individual distributions determined at
a neutrino factory is uniformly
quite low, indicating that a model--independent
flavor and antilabor separation is possible to 10\%--20\% accuracy
in most of the accessible kinematic range.

Such detailed knowledge of the flavor content of the nucleon, besides
providing interesting clues on the nucleon structure, is crucial in
extracting information on possible new physics signal from
experimental data. For example, recently  
a $\sim 3\sigma$ discrepancy has been found between 
the
determination of the Weinberg
angle from the total $\nu$ DIS cross section~\cite{nutevstw} and
 the best fit standard
model prediction. However, the most likely explanation 
of the effect~\cite{DFGRS}
is  a
difference in shape between $\bar s$ and $s$ distributions, which was
disregarded in the analysis of Ref.~\cite{nutevstw}. This $s-\bar s$
difference is
compatible with current uncertainties, and favored by the
analysis of Ref.~\cite{BPZ}, but essentially impossible to determine
accurately with present-day data. This kind of situation is bound to become
increasingly more common as subtle new physics effects will have to be
disentangled from the huge standard background at future hadron
colliders. It is unlikely that any other hadron physics
facility (such as, for
instance, the planned electron-ion collider~\cite{eic}) 
could be competitive with a
neutrino factory for flavor and quark-antiquark separation, and reach the
level of accuracy of Fig.~\ref{pdfec}.

\subsection{Polarized DIS and the proton spin puzzle}
Polarized DIS has recently attracted considerable
attention~\cite{spinrev} because 
of the unexpected smallness of the proton's singlet axial charge $a_0$.
In the naive parton
model the singlet axial charge is the fraction of the nucleon spin
which is carried by quarks. The Zweig rule predicts that it should be
approximately equal to the octet axial charge $a_8$, which differs
from it because of the strange contribution, expected to be small in a
nucleon. The octet charge
can
be determined using SU(3) 
from baryon $\beta$--decay constants: $a_8=0.6\pm 30\%$, so the Zweig rule
leads to expect that the quark spin fraction
is around 60\%. However, the experimental value is compatible with
zero. 

Clearly, this result points to a peculiar role of strangeness in the
nucleon, but the issue is made more subtle by the inclusion of QCD
corrections. Indeed,
beyond leading order the axial charge is given by~\cite{anom}
\beq\label{anomeq}
a_0= \Delta \Sigma-\frac{n_f\as}{2\pi} \Delta G,
\eeq
 where $\Delta
\Sigma=\sum_i(\Delta q_i+\Delta \bar q_i)$ is the scale--invariant
quark spin fraction, and $\Delta G$ is
the gluon spin fraction. The latter, due to the axial anomaly,  gives an
effectively leading--order contribution to $a_0$ Eq.~(\ref{anomeq}). 
On the other hand, gluons
decouple from the octet charge $a_8$. So, 
a first possibility is that gluons are responsible for the difference:
$\Delta G$ is large enough that the
quark spin $\Delta
\Sigma\approx a_8$ even though $a_0<<a_8$.
(`anomaly' scenario). A  
different  option (`instanton' scenario~\cite{inst}) 
is that  $\Delta\Sigma<<a_8$ because
of a large contribution from sea quarks ($\Delta q_i^{\rm sea}=\Delta
\bar q_i^{\rm sea}$)
whose polarization is anticorrelated to that of valence quarks,
possibly because of `instanton' QCD vacuum configurations. Yet
another possibility (`skyrmion' scenario~\cite{skyr}) 
is that $\Delta\Sigma<<a_8$ 
is small because of a
large contribution from `valence' strange quarks $|\Delta s|
>>|\Delta\bar s|$.

At present, flavor separation is only possible 
in the isotriplet sector, and only the total
quark and gluon spin fractions can be
extracted from NC DIS data~\cite{abfr}, and then with modest accuracy: 
$\Delta G(1,1\,{\rm GeV}^2)=
0.8\pm0.2$,
$\Delta \Sigma (1) = 0.38\pm0.03$. It is of course impossible to polarize
the kind of targets which are required for present--day neutrino DIS
experiments, so little
information on strangeness and no information at all 
on the quark--antiquark separation is available in the polarized case.
At a neutrino factory, significant rates could be achieved with small
targets~\cite{RDBnu}: with a detector radius of 50~cm, 100~m length, 
the structure functions $g_1$,  $g_5$ could be
independently measured to an accuracy which is about one order of
magnitude better than that with which $g_1$ is determined 
in present charged lepton DIS experiments.   
On the basis of such data,
the distinct scenarios could be well separated from each
other~\cite{fmr}: for instance
in an `instanton' scenario
 $\left[\Delta s-\Delta\bar s\right](1,1\,{\rm
GeV}^2)=-0.007\pm0.007$; while in a 
`skyrmion''  scenario one would observe $\left[\Delta s-\Delta\bar s\right](1,1\,{\rm
GeV}^2)=-0.106\pm0.008$. In fact
 full flavor separation 
at the level of first moments would be possible. 
Again, the same result
could be hardly achieved anywhere else, since forthcoming experiments are
unlikely to determine $\Delta s$ and $\Delta \bar s$ 
better than current experiment
determined the unpolarized $s$ and $\bar s$.

\section{THE NEUTRINO FACTORY AS A CHARM FACTORY}
\begin{wrapfigure}{l}{0.27\linewidth}\vskip-1.cm
\includegraphics[width=16pc,clip,angle=-90]{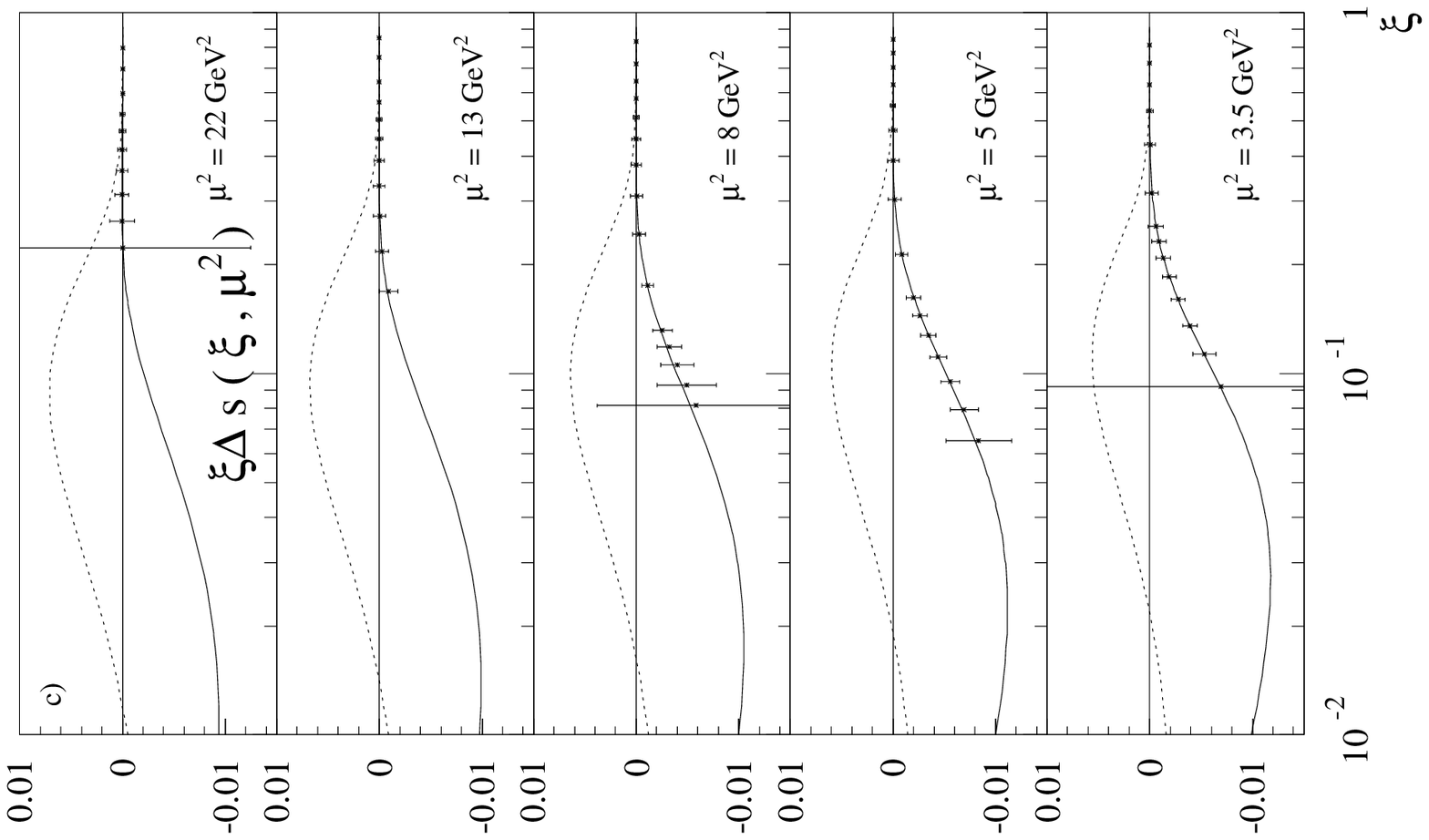}
\vskip-.5cm
\caption{\label{f2c}
Estimated statistical errors on the polarized strange
distribution compared to some current parametrizations.}
\vskip-2.cm
\end{wrapfigure}
Semi-inclusive experiments are a large fraction of the physics program
of present-day electron scattering experiments such as
COMPASS~\cite{compass}  and HERMES~\cite{hermes}. At a neutrino
factory, semi-inclusive charm production would be copious and easy to
detect, because of the pair of opposite-sign muons in the final state
which characterize the charm decay. A precise determination of the unpolarized
and polarized tagged charm structure functions would be possible, since
\beq
\label{cstrf}
F^{\wup}_{1,c} (x,Q^2) = 
|V_{cs}|^2  s (\xi,\mu_c^2) + |V_{cd}|^2  d (\xi,\mu_c^2);
\eeq
\beq
g^{\wup}_{1,c} (x,Q^2) = 
|V_{cs}|^2 \Delta s (\xi,\mu_c^2) + |V_{cd}|^2 \Delta d (\xi,\mu_c^2).
\eeq 
 This would allow a precise direct determination of the strange
distribution (see Fig.~\ref{f2c}), 
though in the polarized case one should worry about
possible large gluon corrections, similarly to Eq.~(\ref{anomeq}).
Also, one could perform accurate studies of 
deep-inelastic charm
production close to threshold, 
which is theoretically quite interesting in
perturbative QCD~\cite{hq} due to the simultaneous
presence of two hard scales ($Q^2$ and
$m_c$).

\begin{figure}[t]
\begin{minipage}[t]{70mm}
\includegraphics[width=15.2pc,clip,angle=-90]{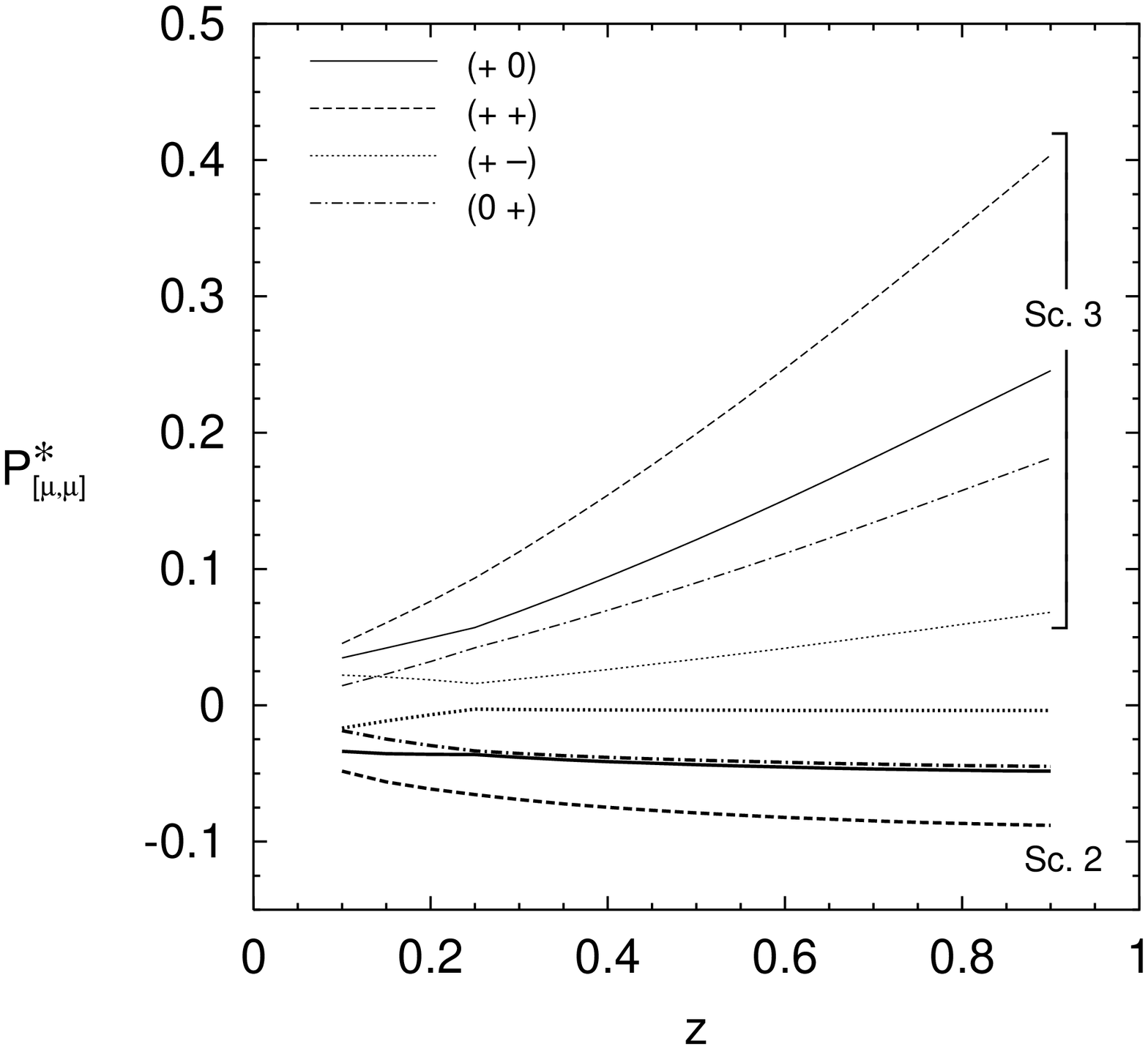}
\end{minipage}
\hspace{\fill}
\begin{minipage}[t]{90mm}
\includegraphics[width=14.9pc,clip,angle=-90]{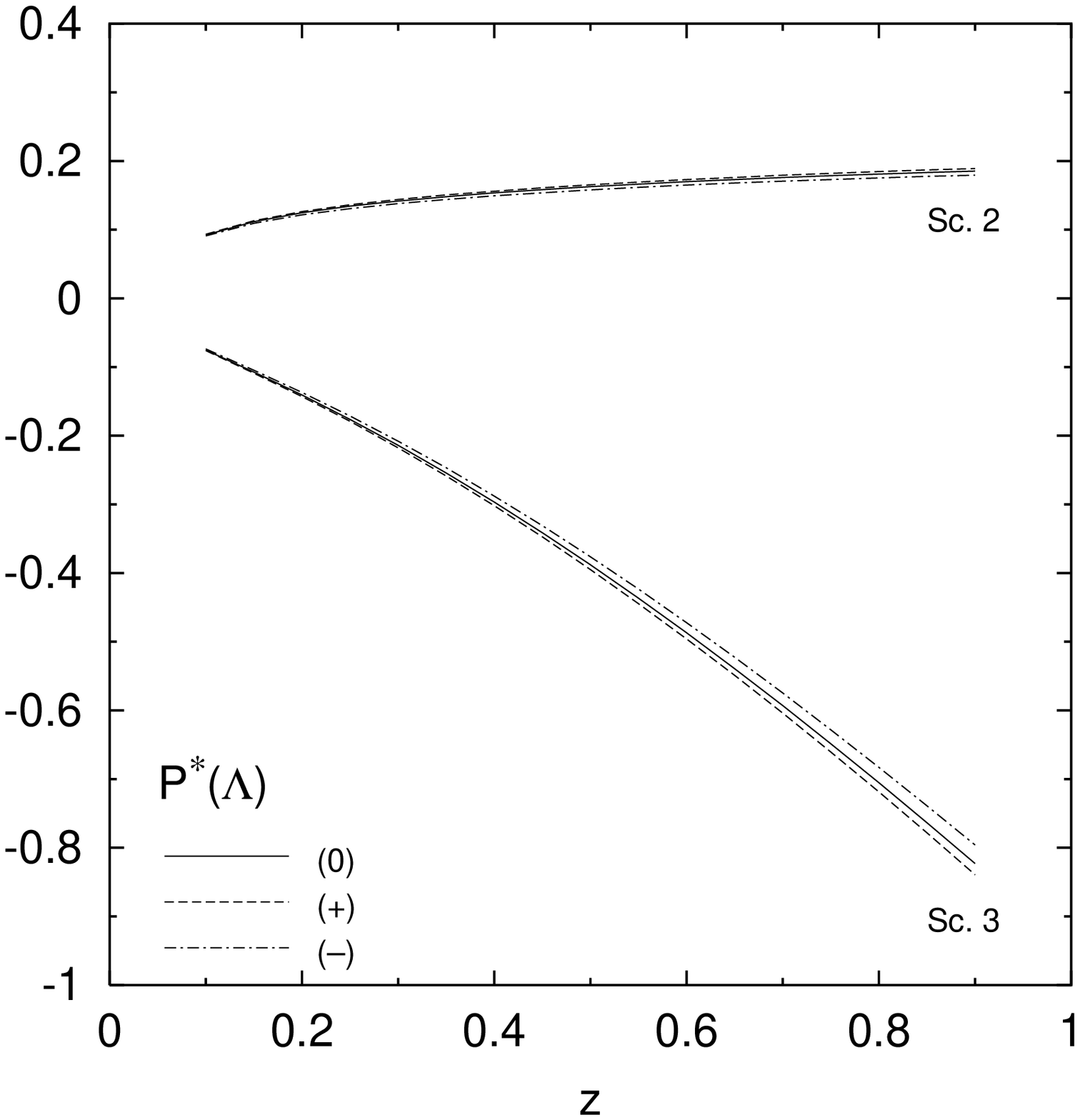}
\end{minipage}
\vskip-.7cm
\caption{\label{lampol}
$\Lambda$ polarization $P(\Lambda)$ at COMPASS (left, from
Ref.~\cite{maurolam}) 
or a
$\nu$ factory (right) for unpolarized (0) or polarized ($\pm$) target
or (COMPASS) beam, in a `proton spin' (Sc.~2) or `SU(3) symmetric'
(Sc.~3) scenario for fragmentation functions.}
\vskip-.6cm
\end{figure}
\section{POLARIZED $\Lambda$ PRODUCTION AND FRAGMENTATION FUNCTIONS}
Semi-inclusive production of specific particles offers information on
both parton distributions and fragmentation functions. An
example is polarized $\Lambda$ production: the $\Lambda$ polarization 
can be determined from the angular distribution of its
decay, and one can construct asymmetries which are directly sensitive
to polarized fragmentation functions. The latter are currently poorly
known, and can only be determined on the basis of theoretical
assumptions~\cite{fragm}: \eg\ that they are SU(3) flavor symmetric, or on the
contrary that the octet combination is much larger than the singlet,
in analogy to what happens for the proton spin fraction. 
Predictions for the asymmetry 
$P(\Lambda)\equiv
\frac{d\sigma^{\Lambda^+}-d\sigma^{\Lambda^-}}{d\sigma^{\Lambda+
\bar\Lambda}}$ as a function of the momentum fraction $z$ carried by
the fragmenting quark are shown in Fig.~\ref{lampol} for fragmentation
functions based on different theoretical assumptions, and for $e^-$
DIS (COMPASS) or $\nu$ DIS. Just as for structure functions, the
flavor dependence of $\nu$ couplings makes results with a $\nu$ beam
significantly more sensitive to the flavor structure of the
fragmenting hadron.

\section{EXCLUSIVE PRODUCTION AND GENERALIZED PDFS}
\begin{wrapfigure}{l}{0.5\linewidth}
\begin{center}
\vskip-1.8cm
\includegraphics[width=0.5\textwidth,clip]{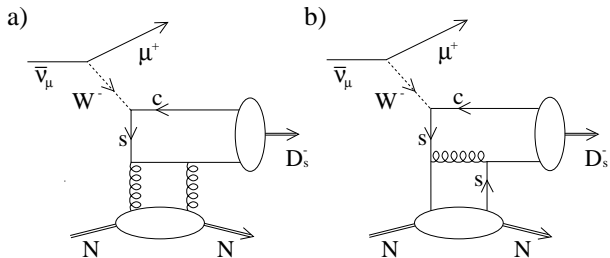} 
\end{center}\vskip-.3cm
\caption{\label{SPD}
Two of the Feynman diagrams  for hard exclusive $D_s$ production.}
\vskip-.7cm
\end{wrapfigure}
Hard QCD factorization can be generalized to several less inclusive
processes by introducing 
generalized parton
distributions~\cite{gpd} (GPD) $F(x,t,Q^2)$. These quantities 
interpolate between the usual nucleon form factor $G(t)$ (which
is related to the first $x$--moment of $F$) and parton distribution
$F(x)$ (related to the $t\to0$ limit of $F$). 
An example of such process is exclusive $D_s$ production (Fig.~\ref{SPD}).
In the kinematic region where the virtuality $Q^2$
of the $W$ boson  is  large compared to the nucleon momentum transfer
$t$ and all masses, the cross section for
the process factorizes as
\beq
\label{gpdfact}
\frac{d\sigma}{d x_{Bj} d Q^2 d t}(W^-+N\to D^-_s+N)=H\otimes
\Phi_D\otimes F,
\eeq
 where
$H$ is the cross--section for the underlying
hard  perturbative parton subprocess,
$\Phi_D$ is the  $D^-$ fragmentation function, and
$F$ is the GPD.
 The estimated cross
section for this process 
is $\sigma=2.2\times
10^{-5}$~pb. 
The  CHORUS collaboration has observed one such event, but in the
region
$Q^2\lsim t\approx1$~GeV$^2$. 
At a neutrino factory, one would observe  $10^4$ event/yr, thus
offering the possibility of a good determination of the GPD.
\section{OUTLOOK}
The focus of studies 
 of the structure of hadrons is currently moving toward either precision
measurements (\eg\ for parton distributions) or rare processes (\eg\
 for generalized parton distributions). Both  are well served by 
the peculiar features of a neutrino beam, namely the availability of a
probe which depends both on spin and flavor.
The physics potential of such a beam can only be exploited
given high enough intensity. However,   a 
high intensity neutrino beam opens the possibility to a class of
 measurements which are essentially impossible at any other
 facility. Even though the time scale for a neutrino factory is
 greater than ten years, it is unlikely that many relevant issues
 of hadron physics, such as the proton spin puzzle, will find a
 satisfactory experimental answer elsewhere.\\
{\bf Acknowledgement:} I thank E.~de~Sanctis for inviting me to
 participate in this
 stimulating meeting.


\vskip-.5cm

\begin{thebibliography}{9}

\bibitem{cernrep}
M.~L.~Mangano {\it et al.},
hep-ph/0105155.
\bibitem{cernmunu}
{\tt http://muonstoragerings.web.cern.ch/muonstoragerings/}.
\bibitem{usrep}
I.~I.~Bigi {\it et al.},
hep-ph/0106177;\\
T.~Adams {\it et al.},
hep-ph/0111030.
\bibitem{status}
M.~M.~Alsharoa {\it et al.},
hep-ex/0207031.
\bibitem{pdfs}
J.~Pumplin, D.~R.~Stump, J.~Huston, H.~L.~Lai, P.~Nadolsky and W.~K.~Tung,
JHEP {\bf 0207} (2002) 012;\\
A.~D.~Martin, R.~G.~Roberts, W.~J.~Stirling and R.~S.~Thorne,
Eur.\ Phys.\ J.\ C {\bf 23} (2002) 73.
\bibitem{alek}
S.~I.~Alekhin,
Phys.\ Rev.\ D {\bf 63} (2001) 094022.
\bibitem{BPZ}
V.~Barone, C.~Pascaud and F.~Zomer,
Eur.\ Phys.\ J.\  {\bf C12}, 243 (2000).
\bibitem{DFGRS}
S.~Davidson, S.~Forte, P.~Gambino, N.~Rius and A.~Strumia,
JHEP {\bf 0202} (2002) 037.
\bibitem{nutevstw}
G.~P.~Zeller {\it et al.}  [NuTeV Collaboration],
Phys.\ Rev.\ Lett.\  {\bf 88} (2002) 091802.
\bibitem{eic}
A.~L.~Deshpande,
Nucl.\ Phys.\ Proc.\ Suppl.\  {\bf 105} (2002) 178;
see also R.~Milner, these proceedings.
\bibitem{spinrev}
See \eg\ S.~Forte,
hep-ph/9409416;
\bibitem{anom}
G.~Altarelli and G.~G.~Ross, \PL\vyp{B212}{1988}{391};
\bibitem{inst}
S.~Forte, \PL\vyp{B224}{1989}{189}; \NP\vyp{B331}{1990}{1}.
\bibitem{skyr}
S.~J.~Brodsky, J.~Ellis and M.~Karliner, \PL \vyp{B206}{1988}{309}.
\bibitem{RDBnu}
R.~D.~Ball, D.~A.~Harris and K.~S.~McFarland,
hep-ph/0009223.
\bibitem{fmr}
S.~Forte, M.~L.~Mangano and G.~Ridolfi,
Nucl. Phys. {\bf B602} (2001) 585.
\bibitem{abfr}
G.~Altarelli, R.~D.~Ball, S.~Forte and
G.~Ridolfi, {\it Acta Phys. Pol.} {\bf B29} (1998) 1145.
\bibitem{compass}
F.~Bradamante  [COMPASS Collaboration],
Nucl.\ Phys.\ A {\bf 622} (1997) 50C.
\bibitem{hermes}
E.~C.~Aschenauer  [HERMES Collaboration],
in ``QCD@WORK'' (AIP, New York 2001); see also K.~Rith, these
proceedings.
\bibitem{hq}
W.~K.~Tung, S.~Kretzer and C.~Schmidt,
J.\ Phys.\ G {\bf 28} (2002) 983.
\bibitem{fragm}
D.~de Florian, M.~Stratmann and W.~Vogelsang,
Phys.\ Rev.\ D {\bf 57} (1998) 5811.
\bibitem{maurolam}
M.~Anselmino, M.~Boglione, U.~D'Alesio and F.~Murgia,
Eur.\ Phys.\ J.\ C {\bf 21} (2001) 501; see also M.~Anselmino, these
proceedings 
\bibitem{gpd}
A.~V.~Radyushkin,
hep-ph/0101225; see also A.~Radyushkin, these proceedings.
\end{thebibliography}
\end{document}